\documentclass[
amsmath, %
floatfix, %
twocolumn, %
reprint, %
prb, %
aps, %
citeautoscript, %
longbibliography, %
final, %
]{revtex4-2}
\renewcommand{\thispagestyle}[1]{}

\usepackage[]{newtxtext}
\usepackage[subscriptcorrection,nosymbolsc,smallerops,bigdelims]{newtxmath}
\DeclareMathAlphabet{\mathcal}{OMS}{cmsy}{m}{n}
\DeclareMathAlphabet{\mathbcal}{OMS}{cmsy}{b}{n}
\usepackage{bm}

\usepackage[utf8]{inputenc}
\usepackage[T1]{fontenc}

\usepackage[]{graphicx}
\usepackage{latexsym}
\usepackage{color}
\usepackage{mathtools}
\usepackage[]{xcolor}
\usepackage{bbold}
\usepackage{multirow}
\usepackage{siunitx}
\usepackage{hyperref}
\hypersetup{
	colorlinks,
	linkcolor={blue!90!black},
	citecolor={blue!90!black},
	urlcolor={blue!90!black}
}
\let\oldeqref\eqref
\renewcommand*{\eqref}[1]{%
	\hyperref[#1]{\oldeqref{#1}}%
}

\usepackage{floatrow}
\newcommand{\mr}[1]{\mathrm{#1}}

\newcommand{\isubfigref}[2]{Figure~\hyperref[#1]{\ref*{#1}(#2)}}

\newcommand{\isubfigsref}[3]{Figures~\hyperref[#1]{\ref*{#1}(#2)}-\hyperref[#1]{\ref*{#1}(#3)}}
\newcommand{\subfigsref}[3]{Figs.~\hyperref[#1]{\ref*{#1}(#2)}-\hyperref[#1]{\ref*{#1}(#3)}}

\mathchardef\mhyphen="2D

\DeclarePairedDelimiter\lr{\lparen}{\rparen}
\DeclarePairedDelimiter\Lr{\lbrack}{\rbrack}

\DeclarePairedDelimiter\abs{\lvert}{\rvert}

\DeclarePairedDelimiterX{\comm}[2]{\lbrack}{\rbrack}{#1, #2}
\DeclarePairedDelimiter\ket{\lvert}{\rangle}
\DeclarePairedDelimiter\bra{\langle}{\rvert}
\DeclarePairedDelimiterX{\braket}[2]{\langle}{\rangle}{#1\delimsize\vert #2}
\DeclarePairedDelimiterX{\ketbra}[2]{\rvert}{\lvert}{#1 \delimsize\rangle\!\delimsize\langle #2}
\DeclarePairedDelimiterX{\matrixel}[3]{\langle}{\rangle}{#1 \delimsize\vert #2 \delimsize\vert #3}

\newcommand{\hc}{\mr{H.c.}}

\newcommand{\eqnref}[1]{Eq.~\eqref{#1}}

\newcommand{\figref}[1]{Fig.~\ref{#1}}
\newcommand{\subfigref}[2]{Fig.~\hyperref[#1]{\ref*{#1}(#2)}}

\usepackage[low-sup]{subdepth}
\usepackage{ragged2e}

\definecolor{cbred}{HTML}{e31a1c}
\definecolor{cbgreen}{HTML}{33a02c}
\definecolor{cbblue}{HTML}{176aa7}
\definecolor{cborange}{HTML}{ff7f00}
\definecolor{cbviolet}{HTML}{6a3d9a}
\definecolor{cbbrown}{HTML}{b15928}

\definecolor{cblred}{HTML}{fb9a99}
\definecolor{cblgreen}{HTML}{b2df8a}
\definecolor{cblblue}{HTML}{a6cee3}
\definecolor{cblorange}{HTML}{fdbf6f}
\definecolor{cblviolet}{HTML}{cab2d6}
\definecolor{cblbrown}{HTML}{ffff99}

\definecolor{CBRed}{RGB}{228,26,28}
\definecolor{CBBlue}{RGB}{55,126,184}
\definecolor{CBGreen}{RGB}{77,175,74}
\definecolor{CBPurple}{RGB}{152,78,163}
\definecolor{CBOrange}{RGB}{255,127,0}
\definecolor{CBYellow}{RGB}{255,255,51}
\definecolor{CBBrown}{RGB}{185,40,15}
\definecolor{CBPink}{RGB}{245,15,180}
\definecolor{CBGrey}{RGB}{153,153,153}

\newcommand{\symRedCircle}{\textcolor{CBRed}{\ensuremath{\bullet}}}
\newcommand{\symBlueSquare}{\textcolor{CBBlue}{\ensuremath{\blacksquare}}}
\newcommand{\symGreenTri}{\textcolor{CBGreen}{\ensuremath{\blacktriangle}}}
\newcommand{\symOrangeStar}{\textcolor{CBOrange}{\ensuremath{\star}}}
\newcommand{\symPurpleDiam}{\textcolor{CBPurple}{\ensuremath{\blacklozenge}}}
\newcommand{\symBrownCross}{\textcolor{CBBrown}{\ensuremath{\times}}}
\newcommand{\symPinkDownTri}{\textcolor{CBPink}{\ensuremath{\blacktriangledown}}}

\newcommand{\afigref}[1]{Figure~\hyperref[#1]{\ref*{#1}}}
\newcommand{\asubfigref}[2]{Figure~\hyperref[#1]{\ref*{#1}(#2)}}
\newcommand{\asubfigsref}[3]{Figures~\hyperref[#1]{\ref*{#1}(#2)}-\hyperref[#1]{\ref*{#1}(#3)}}

\usepackage[activate={true,nocompatibility},final,tracking=alltext,kerning=true,spacing=true,protrusion=true,factor=1080,stretch=7,shrink=7,selected=true ,letterspace=-0]{microtype}
\usepackage{orcidlink}

\newcommand{\SpinSplit}{\varOmega}

\makeatletter
\def\@add@float{%
  \@pageht\ht\@cclv
  \@pagedp\dp\@cclv
  \unvbox\@cclv
  \@next\@currbox\@currlist{%
    \csname @floatselect@sw@\thepagegrid\endcsname\@currbox{%
      \@ifnum{\count\@currbox>\z@}{%
        \advance\@pageht\@pagedp
        \advance\@pageht\vsize
        \advance\@pageht-\pagegoal
        \@addtocurcol
      }{%
        \@addmarginpar
      }%
    }{%
      \@resethfps
      \@ifnum{\pagegrid@cur=\@ne}{%
        \@tempdima\@colht
        \@addtodblcol
        \advance\@tempdima-\@colht
        \@ifdim{\@tempdima>\z@}{%
          \global\advance\@colroom-\@tempdima
        }{}%
      }{%
        \@cons\@deferlist\@currbox
      }%
    }%
  }{%
    \@latexbug
  }%
  \@ifnum{\outputpenalty<\z@}{%
    \@if@sw\if@nobreak\fi{%
      \nobreak
    }{%
      \addpenalty\interlinepenalty
    }%
  }{}%
  \set@vsize
}
\makeatother

\begin{document}

\title{Fast universal parametric spin control in an acoustically modulated quantum dot}

\author{Mateusz Kuniej\,\orcidlink{0000-0001-5476-4856}}
\affiliation{Institute of Theoretical Physics, Wroc\l{}aw University of Science and Technology, 50-370 Wroc\l{}aw, Poland}

\author{Micha{\l} Gawe{\l}czyk\,\orcidlink{0000-0003-2299-140X}}
\affiliation{Institute of Theoretical Physics, Wroc\l{}aw University of Science and Technology, 50-370 Wroc\l{}aw, Poland}

\begin{abstract}
Quantum communication, distributed computing, and hybrid architectures rely on nodes enabling coherent control of qubits and coupling to propagating quantum modes. While semiconductor quantum-dot (QD) spins couple to microwave and optical photons, weak interaction with mechanical waves has limited the integration of single-QD spin qubits into on-chip, acoustically coupled hybrid systems. The existing theory of acoustic QD spin control suffers from a limited range of rotation-axis angles, enforcing complex realizations of gate primitives and long gate times, leaving little margin against decoherence from trion decay and quasi-static nuclear-spin noise. We propose parametric control that overcomes these problems. We use far-detuned optical coupling to a trion state to dress and thus mix spin states, combined with acoustic modulation of the optical transition energy. Instead of relying on direct acoustic resonance with the spin splitting that leads to significant bottlenecks, we induce spin rotations parametrically via resonance with the dressed-spin splitting. We thus develop a spin analog of the ``swing-up'' charge-state excitation. Our scheme provides fast universal qubit control with nearly arbitrary rotation axes. Our ${\sim}$155~ps Pauli-$X$ gate duration is ${\sim}290\times$ faster than in the previous acousto-optical formulation and ${\sim}14\times$ faster than optical Faraday-geometry spin rotation. The parametric scheme naturally enables higher-harmonic processes. Numerical simulations for ${\sim}44$~GHz acoustic driving show average gate fidelity $\ge99.9\%$ even for uncooled nuclear-spin environments of GaAs and InAs QDs for trion lifetime $\gtrsim1.25$~ns. These metrics suggest practically usable control and may introduce a spin-phonon interface with high interaction rates, versatility, and multi-phonon processes, essential for future acoustically coupled hybrid architectures.
\end{abstract}

\maketitle

\section*{Introduction}

Coherent control of solid-state spins and their coupling to propagating quantum modes are two essential requirements for building scalable quantum nodes for processing and exchange of quantum information~\cite{SchimpfPRX2025, Awschalom2018, Appel2025}. Acoustic fields are particularly promising in this context as they can serve both as a local control field and an on-chip quantum bus connecting confined spin states to a broad range of solid-state degrees of freedom in quantum hybrid systems~\cite{Kurizki2015, Schuetz2015, Krenner2026, Clerk2020}. However, to realize this potential, acoustic spin manipulation must be universal and robust, and provide gate speeds that are competitive with established control techniques.

Semiconductor quantum dots (QDs) are widely studied as quantum light emitters that stand out for the quality of single photons~\cite{Musial2020, Tomm2021}, their entangled pairs~\cite{Liu2019, Vajner2022}, and ability to generate cluster states~\cite{Lindner2009, Economou2010, Schwartz2016}. QD spin states have attracted attention due to their relatively long coherence times~\cite{NguyenPRL2023}, the controllable interface with optical photons~\cite{SchimpfPRX2025}, and the possibility of harnessing~\cite{SunPRL2012, EthierMajcherPRL2017, JacksonPRX2022, NguyenPRL2023} and coherently controlling their nuclear spin environment~\cite{Shofer2025}, and turning it into a quantum register~\cite{Chekhovich2020, Appel2025}. Thus, QDs are a promising platform that meets both of the aforementioned requirements for a quantum node, currently exploiting photons and enabling long-distance state transfer.

Due to low sound velocity in solids, acoustic waves and their quantized modes occupy the short-wavelength regime, with wavelengths several orders of magnitude shorter than electromagnetic waves at comparable frequencies, which is useful for extending the use of QD spins toward compact, on-chip, short-range communication and control. This acoustically mediated integration is enabled by strain and piezoelectric couplings to a broad range of systems in solids~\cite{Krenner2026}, including superconducting qubits, a leading platform in current quantum computing~\cite{Gustafsson2014, Manenti2017, Li2025}, thus facilitating phonon-mediated heterogeneous information processing and transduction between otherwise disparate quantum degrees of freedom~\cite{Schuetz2015}. While proposals for realizing acoustically driven control and state transfer involving double-QD singlet-triplet qubits exist~\cite{Schuetz2015, Guo2025}, acoustically driven control of individual confined spins would provide the missing link for a mechanical interface to QD quantum nodes.

Realizing such an interface has been hindered by the weak intrinsic coupling between lattice deformation and spin in high-symmetry solid-state systems~\cite{Golovach2004, MielnikPyszczorski2018}, where the dominant deformation-induced interactions mostly preserve the spin projection. One route towards overcoming this limitation was taken by exploiting virtual optical coupling to a trion state in a $\Lambda$-system that breaks spin conservation in the ground-state doublet, enabling coherent acoustic spin rotation~\cite{KuniejPRL2026}.
However, the resonant character of the method, where the acoustic frequency must match the Zeeman splitting, combined with relatively low available acoustic frequencies, results in long gate operation times that exceed the involved trion lifetime and inhomogeneous spin dephasing time, leading to strong decoherence constraints. Moreover, the restricted set of feasible qubit rotation axes necessitates composite pulse sequences even for simple spin gates and the development of custom noise-mitigating sequences. These limitations affect not only the acoustic spin gates, but also the usefulness of the same interaction as a coherent spin-phonon interface for state transfer and hybrid quantum operations. Thus, further exploration of QD acousto-spintronics requires overcoming bottlenecks of gate operation speed and lack of simple universal control primitives, which, in turn, translate into a spin-phonon interface restricted in both speed and functionality.

Here, we overcome these obstacles by proposing a parametric acousto-optical method for spin control. We not only rely on virtual optical coupling to the trion to enable acoustic spin transitions but also exploit optical dressing to move away from the bare-spin resonance. Instead, to achieve coherent manipulation, we use the parametric resonance that occurs when the acoustic frequency matches the splitting between optically dressed spin states, in analogy to the ``swing-up'' method for preparation of charge states in QDs~\cite{Bracht2021, Karli2022, Bracht2023, Boos2024, Kuniej2025} and the recently observed hybrid acousto-optical double dressing of a two-level quantum emitter~\cite{ZhanPRL2026}. Thanks to the parametric character of the control, the acoustic frequency no longer needs to be resonant with the Zeeman splitting. Since the dressed spin splitting is itself modulated, higher-harmonic processes arise naturally and can also be utilized~\cite{Kuniej2026Arxiv}. We show that this approach is substantially faster: the ${\sim}155$~ps $\pi$ rotation time (defined as $4\sigma$ of the Gaussian control pulse) is ${\sim}290 \times$ shorter than the noise-mitigating gate reported in Ref.~\onlinecite{KuniejPRL2026} and significantly shorter even compared to all-optical Faraday control with few-nanosecond timescales~\cite{Koong2026}. This substantial improvement stems from relaxed restrictions on acoustic field parameters. In contrast to Ref.~\onlinecite{KuniejPRL2026}, the present scheme does not require the acoustic amplitude and frequency to be comparable, and allows one to compensate for weak acoustic driving by applying a stronger optical pulse without compromising the desired qubit rotation axis.
Thanks to reduced gate times, we achieve 99.9\% fidelity even for noise amplitudes of uncooled thermal nuclear-spin environments of typical QDs for the fundamental acoustic harmonic and trion radiative lifetime above $1.25$~ns, and 99.8\% fidelity for the second-harmonic process for representative cases of cooled environments and trion lifetime $\gtrsim 2.5$~ns. For a future spin-phonon interface, these advancements have the potential to translate into higher interaction rates, greater interface control, and availability of multiphonon transitions. Thus, the approach may not only open the way for acoustic QD spin control, but also enable coherent spin-phonon state transfer and QD integration into hybrid quantum architectures.

\section*{Theory}\label{sec:systemQD}

\subsection*{System and its model}

\begin{figure*}[!t]
    \centering
    \includegraphics[width=1\textwidth]{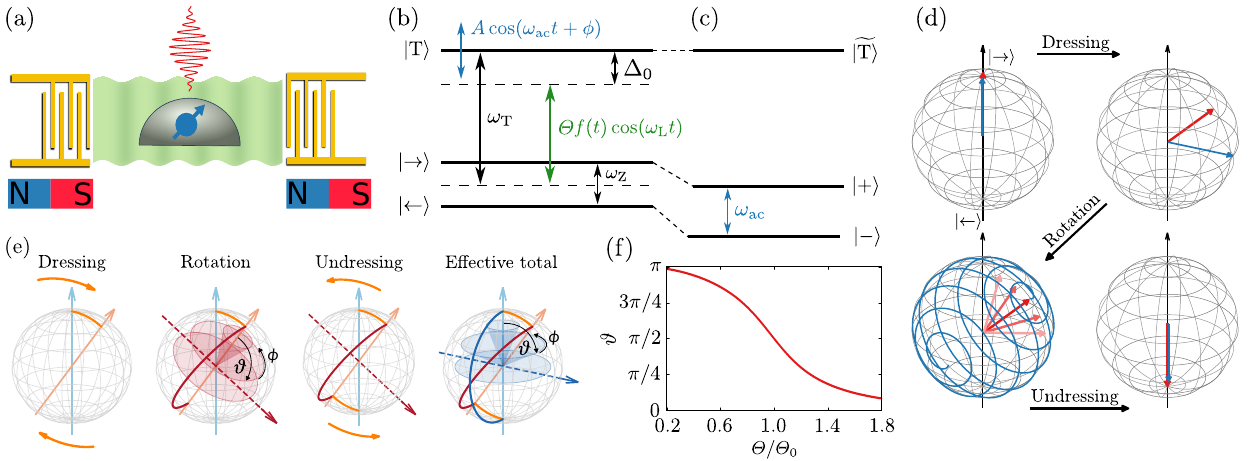}
    \caption{Acousto-optical parametric spin control scheme.
    (a) Principle of the method: a single electron trapped in a QD with a magnetic field perpendicular to the optical axis (Voigt) is modulated with a monochromatic surface acoustic wave.
    (b) Energy diagram with external fields marked. Zeeman doublet $\ket{\rightarrow}$, $\ket{\leftarrow}$ is coupled off-resonantly to the trion state $\ket{\mathrm{T}}$ by a laser (green arrow). The trion state energy is modulated by an acoustic wave (blue arrow).
    (c) Optically dressed states. The acoustic field frequency (blue line) is tuned to drive the $\ket{+}\leftrightarrow\ket{-}$ dressed spin transition.
    (d) Spin rotation: exact evolution in the lab frame. The electron spin (blue arrow) is initially in the $\ket{\rightarrow}$ state. Without optical and acoustic fields, the rotation axis (red arrow) points to the pole. Next, the laser pulse couples both spin states via $\ket{\mathrm{T}}$, simultaneously tilting the rotation axis. The acoustic field periodically modulates this axis, leading to spiral evolution on the Bloch sphere (blue line).
    (e) Spin rotation: effective evolution in the dressed qubit rotating frame: dressing (orange), rotation (red; cones of possible rotation axes shown), undressing (orange), and the total effective rotation (blue).
    (f) The range of available inclinations of the spin rotation axis according to Eq.~\eqref{eq:hamiltonianTimeIndependent} for parameters from the third column of Table~\ref{tab:parameters} and varied $\varTheta$.
    }
    \label{fig:systemQD}
\end{figure*}

\afigref{fig:systemQD} shows schematically the idea presented here. \asubfigref{fig:systemQD}{a} illustrates the system and idea of spin control in an acoustically modulated QD. Spin states are split in a Voigt-configuration magnetic field and are optically dressed by a laser pulse. A continuous-wave coherent acoustic field generated, e.g., by an interdigital transducer, drives spin rotation while the system is optically dressed. In \subfigref{fig:systemQD}{b} we show the energy diagram of the system. The QD electron spin states $\ket{\rightarrow}$ and $\ket{\leftarrow}$ are split by the Zeeman energy $\hbar\omega_{\mathrm{Z}}$ and coupled off-resonantly to a trion (negatively charged exciton) state via a pulsed $\sigma^+$-polarized laser. Selection rules in typical semiconductor QDs ensure that coupling of both spin states to one of the trion states is equal in this configuration~\cite{SchimpfPRX2025}, while in the far-detuned case the other trion state can be safely neglected (see Appendix~\ref{app:three-level}). The laser energy $\hbar\omega_{\mathrm{L}}$ is detuned from the trion optical transition energy $\hbar\omega_{\mathrm{T}}$ by $\Delta_0 = \omega_{\mathrm{T}} - \omega_{\mathrm{L}}$ to avoid populating the trion, which would cause decoherence, and only exploit it virtually. Crucially, the system is modulated acoustically by, e.g., a surface acoustic wave (SAW), which introduces periodic changes to the detuning of the optical transition~\cite{Wigger2021, Weiss2021, KuniejPRL2026}. \asubfigref{fig:systemQD}{c} shows a dressed-state picture of the system. In this symmetric $\Lambda$-like system, the optical dressing effectively creates opposite-spin admixtures in the spin states and consequently relaxes spin conservation, while the detuned trion dressed state $\ket{\widetilde{\mathrm{T}}} \approx \ket{\mathrm{T}}$ almost does not mix with the Zeeman doublet. When the acoustic field frequency $\omega_{\mathrm{ac}}$ is tuned to the dressed-state spin splitting, coherent rotations are enabled. In \subfigref{fig:systemQD}{d}, we sketch the evolution of the spin qubit on a Bloch sphere in the lab frame. The spin is initialized in the $\ket{\rightarrow}$ state. Next, the states become dressed during the pulse turn-on, and, subsequently, the acoustic field induces spin rotation in the presence of optical dressing. During the quasi-adiabatic optical pulse switch-off, the system undresses to the desired final state.

The full Hamiltonian of the system coupled to external fields is given by
\begin{equation}
    H(t) = \!\!\!\!\sum_{i\in\{\rightarrow, \leftarrow, \mathrm{T}\}}\!\!\!\!\hbar\omega_{i}\ketbra{i}{i} - \bm{d}\cdot\bm{E}(t) + H_{\mathrm{ac}}(t),
\end{equation}
where $\bm{E}(t) = \bm{E}_0(t)\cos(\omega_{\mathrm{L}}t)$ is the classical optical field and $\bm{d}$ is the dipole moment operator describing the light-matter coupling in the dipole approximation, for which we assume $\bra{i} \bm{d}\ket{i} = 0$. Furthermore, $H_{\mathrm{ac}}(t) = \hbar A\cos(\omega_{\mathrm{ac}}t + \phi)\ketbra{\mathrm{T}}{\mathrm{T}}$ is the acoustic modulation of the $\ket{\mathrm{T}}$ state with a constant amplitude $A$, angular frequency $\omega_{\mathrm{ac}}$, and phase $\phi$. In the frame rotating at $\omega_{\mathrm{L}}$ and in the rotating wave approximation, the Hamiltonian reads
\begin{equation}
    \begin{split}
        H(t) = {}& \frac{1}{2}\hbar\omega_{\mathrm{Z}}\left(\ketbra{\rightarrow}{\rightarrow} - \ketbra{\leftarrow}{\leftarrow}\right) + \hbar\Delta(t)\ketbra{\mathrm{T}}{\mathrm{T}} \\
        &+\frac{1}{2}\hbar\varTheta f(t)\left(\ketbra{\rightarrow}{\mathrm{T}} + \ketbra{\leftarrow}{\mathrm{T}} + \hc\right),
    \end{split}
    \label{eq:spinQDHamiltonian}
\end{equation}
where 
\begin{equation}
    \begin{split}
        \Delta(t) = \Delta_0 + \frac{1}{\hbar}\bra{\mathrm{T}}H_{\mathrm{ac}}(t)\ket{\mathrm{T}}
        = \Delta_0 + A\cos(\omega_{\mathrm{ac}}t + \phi),
    \end{split}
\end{equation}
is the acoustically modulated instantaneous detuning, $\varTheta$ is the area of the optical pulse, and $f(t)$ is its envelope normalized to unit area. To avoid accidental resonances when the amplitude of the acoustic field matches the detuning, and to keep the trion occupation low, we consider $|\Delta_0|>A$. 

\subsection*{Effective Hamiltonian}
We aim to achieve full spin control by combining far-detuned optical coupling with acoustic modulation at a frequency matching the splitting between optically dressed spin states. We expect this to induce an acousto-optical evolution similar to ``swing-up'' excitation of charge states, resulting in acoustic rotation of the dressed spin state. As shown below, the dressing amounts to a rigid rotation of the Bloch sphere, so a rotation of the dressed spin translates directly into a corresponding rotation of the bare spin. This process is illustrated schematically in \subfigref{fig:systemQD}{e} for an $X$ spin gate, where the (un)dressing (orange) and dressed rotation (red) comprise the effective rotation of the bare spin (blue).  

In the next section, we solve the evolution numerically using the full three-level Hamiltonian, but to gain understanding and make predictions, it is convenient to construct an effective model. To do this, we would like to switch to the basis of optically dressed states, which requires diagonalizing the Hamiltonian~\eqref{eq:spinQDHamiltonian}, and then adiabatically eliminate the trion state. Doing it in this intuitive order would, however, spoil the handling. Thus, we reverse the order of the operations and start by eliminating the trion state using adiabatic elimination~\cite{Haken1975, Brion2007}, which is formally valid for small $\varTheta f(t)/(2\abs{\Delta(t)}-\abs{\omega_{\mathrm{Z}}})$~\cite{KuniejPRL2026}. The effective two-level Hamiltonian reads
\begin{equation}\label{eq:eff-H-initial}
    H_{\mathrm{eff}}(t) = \frac{1}{2}\hbar\omega_{\mathrm{Z}}\sigma_{\mathrm{z}} - \frac{\hbar\varTheta^2f^2(t)}{4\Delta(t)}\sigma_{\mathrm{x}},
\end{equation}
where $\sigma_i$ are Pauli matrices in the $\{\ket{\rightarrow}, \ket{\leftarrow}\}$ basis, and we dropped a term proportional to the identity.

To obtain a transparent analytical description of the parametric resonance, we construct a leading-order theory in the acoustic modulation depth $A/\Delta_0$. Expanding the modulated detuning into a series
\begin{equation}
    \frac{1}{\Delta(t)} \approx \frac{1}{\Delta_0}\left[1 - \frac{A}{\Delta_0}\cos(\omega_{\mathrm{ac}}t + \phi)\right],
\end{equation}
allows us to separate the purely optical contribution from acousto-optical effects. The separation will allow analysis in the optical dressed-state language, while retaining only the single $(A/\Delta_0)$-linear monochromatic term enables analytical treatment. Our numerical simulations below use the full three-level Hamiltonian and do not rely on this expansion. In Appendix~\ref{app:nonperturbative}, we present a nonperturbative derivation preserving the full nonlinearity arising from the modulated detuning, and show that it reduces to the leading-order theory given here for $A/\Delta_0\ll1$. Even beyond this limit, the full theory gives structurally identical results with renormalized parameters such as qubit effective drive strength or detuning.

We are now able to distinguish terms generated by purely optical and acousto-optical mechanisms: the quasi-static coupling generated by the pulsed optical field
\begin{equation}
    \omega_{\mathrm{s}}(t) = \frac{\varTheta^2f^2(t)}{2\Delta_0},
    \label{eq:staticShift}
\end{equation}
and the leading-order dynamical coupling that arises due to the acoustic modulation
\begin{equation}
    \omega_{\mathrm{d}}(t) = \omega_{\mathrm{s}}(t)\frac{A}{\Delta_0}\cos(\omega_{\mathrm{ac}}t + \phi).
\end{equation}
With this, we can rewrite the effective Hamiltonian as
\begin{equation}
    \begin{split}
        H_{\mathrm{eff}}(t) \approx
         H_{\mathrm{eff}}^{(0)}(t) + H_{\mathrm{eff}}^{(\mathrm{ac})}(t), 
    \end{split}
\end{equation}
where $H_{\mathrm{eff}}^{(0)}(t) = \hbar\omega_{\mathrm{Z}}\,\sigma_{\mathrm{z}}/2 -\hbar\omega_{\mathrm{s}}(t)\,\sigma_{\mathrm{x}}/2$ and $H_{\mathrm{eff}}^{(\mathrm{ac})}(t) = \hbar\omega_{\mathrm{d}}(t)\sigma_{\mathrm{x}}/2$. One can notice that $\hbar\omega_{\mathrm{d}}(t)$ can be interpreted as a second-order acousto-optical coupling between spin states, and since the acoustic modulation term is no longer diagonal, it drives transitions when $\omega_{\mathrm{ac}} = \omega_{\mathrm{Z}}$ as previously shown in Ref.~\onlinecite{KuniejPRL2026}.

We do not focus here on those known direct resonances. Instead, we continue analyzing the effective Hamiltonian in the optical dressed-state basis. For a slowly-varying laser-pulse envelope, we diagonalize the purely optical quasi-static part $H^{(0)}_{\mathrm{eff}}(t)$ at each time point and find new instantaneous states dressed by light
\begin{equation}\label{eqs:dressedStates}
    \ket{+(t)} = D(\alpha(t)) \, \ket{\rightarrow},~~\ket{-(t)}
    = D(\alpha(t)) \, \ket{\leftarrow},
\end{equation}
with
\begin{equation}
    D(\alpha)=e^{-i\alpha\sigma_y/2}
\end{equation}
describing a simple rigid rotation of the Bloch sphere by the mixing angle $\alpha(t) = -\tan^{-1}(\omega_{\mathrm{s}}(t)/\omega_{\mathrm{Z}})$. The dressed-state spin splitting is $\SpinSplit(t)=\sqrt{\cramped{\omega_{\mathrm{Z}}^2 + \omega^2_{\mathrm{s}}(t)}}$. Nonadiabatic effects are negligible, as the nonadiabaticity parameter remains $\abs{\dot{\alpha}(t)/[2\SpinSplit(t)]}<0.03$ for parameter sets further used and collected in Table~\ref{tab:parameters}.
Since typically $\abs{\omega_{\mathrm{s}}(t)}<\abs{\omega_{\mathrm{Z}}}$, the needed acoustic frequency for the resonance is greater, but of the same order of magnitude as $\omega_{Z}$. We can rewrite $H^{(\mathrm{ac})}_{\mathrm{eff}}(t)$ in the dressed state basis as
\begin{equation}\label{eq:effectiveAcousticDressedStates}
    \widetilde{H}_{\mathrm{eff}}^{\mathrm{(ac)}}(t) = \frac{\hbar\omega_{\mathrm{d}}(t)}{2\SpinSplit(t)}\left[-\omega_{\mathrm{s}}(t)\,\widetilde{\sigma}_{\mathrm{z}} + \omega_{\mathrm{Z}}\,\widetilde{\sigma}_{\mathrm{x}}\right],
\end{equation}
where $\widetilde{\sigma}_i=D(\alpha)\sigma_i D^\dagger(\alpha)$ are the Pauli matrices in the new basis of states. Next, we perform a unitary transformation given by $e^{iS(t)}$, where
\begin{equation}
    S(t) = - \int_{0}^{t}\mathrm{d}\tau \frac{\omega_{\mathrm{d}}(\tau)\,\omega_{\mathrm{s}}(\tau)}{2\SpinSplit(\tau)}\,\widetilde{\sigma}_{\mathrm{z}},
\end{equation}
to remove the $\widetilde{\sigma}_{\mathrm{z}}$ component from $\widetilde{H}_{\mathrm{eff}}^{\mathrm{(ac)}}(t)$, thus moving to the frame in which the acoustic drive is purely transverse. Assuming that the optical pulse is rectangular-like and $\SpinSplit$ and $\omega_{\mathrm{s}}$ are piecewise constant, we can calculate this integral analytically:
\begin{equation}\label{eq:unitaryTransformation}
    S(t) = - \frac{1}{2}\mathcal{A} \Lr*{ \sin(\omega_{\mathrm{ac}}t + \phi) -\sin\phi }\,\widetilde{\sigma}_{\mathrm{z}},
\end{equation}
where we defined $\mathcal{A} = A\omega_{\mathrm{s}}^2/\SpinSplit\Delta_0\omega_{\mathrm{ac}}$ for convenience. Using the Jacobi-Anger formula $e^{-iy\sin(x)}=\sum_{m}J_m(y)e^{-imx}$, followed by a secular approximation to maintain only the $m=n$ phonon processes, gives the following effective Hamiltonian with a transverse acoustic drive
\begin{align}\label{eq:effectiveHamiltonian}
        \widetilde{\mathcal{H}}_{\mathrm{eff}}(t) = \frac{1}{2}\hbar\SpinSplit\,\widetilde{\sigma}_{\mathrm{z}}
            + \hbar g_n \big[ & \!\cos(n\omega_{\mathrm{ac}}t + \varPhi_n(\phi))\,\widetilde{\sigma}_{\mathrm{x}} \notag \\
        + & \sin(n\omega_{\mathrm{ac}}t + \varPhi_n(\phi))\,\widetilde{\sigma}_{\mathrm{y}} \big],\!\!\!\!
\end{align}
where $g_n = n\omega_{\mathrm{ac}}\omega_{\mathrm{Z}}J_n(\mathcal{A})/(2\omega_{\mathrm{s}})$, and $\varPhi_n(\phi) = n\phi -\mathcal{A}\sin\phi$.
This approximation is justified by well-separated multiphonon resonances compared to the driving strength, with $\max_{m\neq n} \abs{ g_m/[(n-m)\omega_{\mathrm{ac}}] }< 0.08$ for all parameter sets in Table~\ref{tab:parameters}.
To eliminate direct time dependence, we perform a final transformation to the rotating frame, with respect to the acoustic frequency, $e^{i n\omega_{\mathrm{ac}}t\widetilde{\sigma}_{\mathrm{z}}/2}$. The time-independent effective Hamiltonian then reads
\begin{equation}\label{eq:hamiltonianTimeIndependent}
    \widetilde{\mathcal{H}}_{\mathrm{eff},n}^{(\mathrm{rf})} = \frac{1}{2} \hbar \bm{\varOmega}_{n}\cdot\widetilde{\bm{\sigma}}
        = D(\alpha) \Lr*{ \frac{1}{2} \hbar \bm{\varOmega}_{n}\cdot\bm{\sigma} } D^\dagger(\alpha),
\end{equation}
where $\bm{\varOmega}_{n} = (2 g_n \cos\varPhi_n(\phi), 2 g_n \sin\varPhi_n(\phi), \delta_n)$ is the instantaneous rotation vector in the rotating frame (fixed during a pulse plateau). Thus, for fixed effective detuning $\delta_n = \SpinSplit-n\omega_{\mathrm{ac}}$ and effective drive strength $g_n \simeq n\omega_{\mathrm{ac}}\omega_{\mathrm{Z}}\mathcal{A}^n/(2^{n+1}n!\,\omega_{\mathrm{s}})$ (we generally consider $\mathcal{A}\ll1$), varying acoustic phase $\phi$ defines, through $\varPhi_n(\phi)\simeq n\phi$, a cone of available rotation axes around the $z$ axis, with the cone angle $\vartheta = \mathrm{atan2}(2 g_n,\delta_n)$, which can be set by detuning. The transformation $D(\alpha)$ rotates the dressed frame with respect to the bare one. In the adiabatic limit, the (un)dressing ramps can be written as $U_{\mathrm{d}}\simeq D P_{\mathrm{d}}$ and $U_{\mathrm{u}}\simeq P_{\mathrm{u}}D^\dagger$, where $P_{\mathrm{d(u)}}$ describes the phase accumulated during (un)dressing. Therefore, the complete bare-to-bare evolution for a rectangular pulse in the $n\omega_{\mathrm{ac}}$ rotating frame is
\begin{align}\label{eq:evolution-effective}
    U_{\mathrm{gate}}
        = U_{\mathrm{u}}e^{-\frac{i}{2} \bm{\varOmega}_{n}\cdot\widetilde{\bm{\sigma}}T} U_{\mathrm{d}} 
        = P_{\mathrm{u}} e^{-\frac{i}{2} \bm{\varOmega}_{n}\cdot\bm{\sigma}T} P_{\mathrm{d}},
\end{align}
with $T$ the rotation time, and describes qubit rotation around the same vector $\bm{\varOmega}_{n}$ in the bare state basis, sandwiched between additional phase rotations accumulated during (un)dressing that can be incorporated into control phases.
For a nonrectangular pulse, there is no single rotating frame in which the effective Hamiltonian is time-independent, and adopting the standard experimental $\omega_{\mathrm{Z}}$ frame is the natural choice as it preserves the stationary logical qubit basis, while residual nonsecular motion during the pulse will be present.
\asubfigref{fig:systemQD}{e} schematically shows the effective evolution together with cones of possible rotation axes, and \subfigref{fig:systemQD}{f} shows the range of available rotation-axis inclinations for parameters further used in the paper (third column in Table~\ref{tab:parameters}) and a laser pulse area varied in the range of $0.2\varTheta_0$--$1.8\varTheta_0$, where $\varTheta_0$ corresponds to $\delta_n=0$.
The scheme allows virtually any rotation axis, more than sufficient for full control over the spin qubit evolution.

After gaining intuition from the effective model, we can confront it with the results of direct numerical simulations. To study the evolution of the closed system, in the next section, we numerically solve the time-dependent Schr{\"o}dinger equation for the Hamiltonian from Eq.~\eqref{eq:spinQDHamiltonian}
\begin{equation}
    i\hbar\ket{\dot{\mathit{\Psi}}(t)} = H(t)\ket{\mathit{\Psi}(t)},  
\end{equation}
where $\ket{\mathit{\Psi}(t)}$ is the state vector of the three-level system, whereas for the open system dynamics, we use the Gorini-Kossakowski-Sudarshan-Lindblad master equation
\begin{equation}
    \mkern-10mu\Dot{\rho}(t) \mkern-1mu=\mkern-1mu {\frac{1}{i\hbar}}\!\left[H(t), \rho(t)\right] {+} \sum_{i} \! \left(L_{i}\rho(t)L^{\dagger}_{i} {-} \frac{1}{2}\left\{L^{\dagger}_{i}L_{i}, \rho(t)\right\} \right)\!,\!
    \label{eq:Lindblad}
\end{equation}
where $\rho(t)$ is the density matrix of the system, $L_{i} = \ketbra{i}{\mathrm{T}}/\sqrt{2\tau}$ are the jump operators for $i = \rightarrow$, $\leftarrow$, and $\tau$ is the trion radiative lifetime. 

\section*{Results}

\subsection*{Efficient $\pi$-rotations}

To study and fully determine the conditions for the desired qubit evolution, we numerically investigate the system in this section. To relate to experimental conditions, we assume the optical pulse envelope to be a normalized Gaussian with a standard deviation $\sigma_{\mathrm{L}}$. Since such realistic pulses do not produce a constant dressed-state splitting, parameters must be optimized. We chose to study the behavior of the system for a given $\hbar\omega_{\mathrm{Z}}$, connected with the time duration of the optical pulse $\sigma_{\mathrm{L}} > 2\pi/\omega_{\mathrm{Z}}$, so that the transitions are performed quasi-adiabatically~\cite{Bracht2021, Bracht2023, Kuniej2025}.

\begin{figure}[tb!]
    \centering
    \includegraphics[width=1\linewidth]{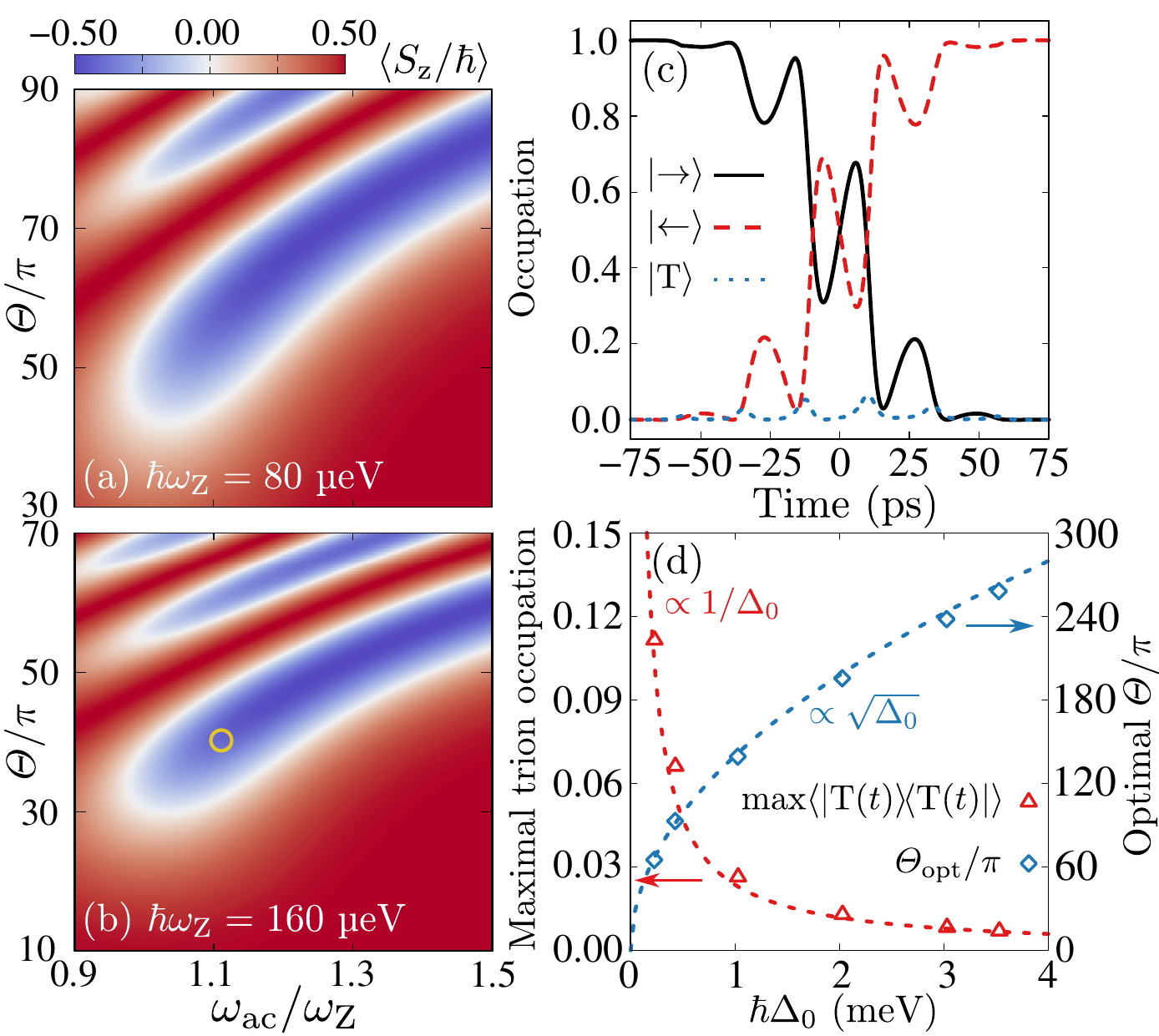}
    \caption{(a) Final value of the $\langle S_{\mathrm{z}}\rangle$ as a function of the optical pulse area and acoustic frequency for $\hbar\omega_{\mathrm{Z}} = 80$~\textmu{}eV, $\hbar\Delta_0 \approx 3.94$~meV, $\hbar A = 2$~meV, and $\sigma_{\mathrm{L}} \approx 78.6$~ps. (b) The same as (a) but for $\hbar\omega_{\mathrm{Z}} = 160$~\textmu{}eV, $\hbar\Delta_0 \approx 3.44$~meV, $\hbar A = 2$~meV, and $\sigma_{\mathrm{L}} \approx 38.8$~ps. The yellow circle shows the optimal parameters for the $\pi$ rotation: $\varTheta\approx38.4\pi$, $\omega_{\mathrm{ac}}\approx1.134\omega_{\mathrm{Z}}$. (c) Time evolution of the three-level system for optimal parameters from (b); black solid and red dashed lines present the occupation of the spin states, while the blue dotted line corresponds to the trion. (d) Maximum trion occupation $\mathrm{max}\langle\ketbra{\mathrm{T}(t)}{{\mathrm{T}(t)}}\rangle$ (red triangles, left axis) and optimized optical pulse area $\varTheta_{\mathrm{opt}}$ (blue squares, right axis) as a function of $\hbar\Delta_0$ for $A = \Delta_0/4$, $\phi = 0$, and $\tau = \infty$. The value of $\mathrm{max}\langle\ketbra{\mathrm{T}(t)}{{\mathrm{T}(t)}}\rangle$ was fitted with a $\sim1/\Delta_0$ function, and $\varTheta_{\mathrm{opt}}$ via square root of the detuning.}
    \label{fig:map_evolution}
\end{figure}

In \subfigsref{fig:map_evolution}{a}{b} we show the average value of $S_{\mathrm{z}} = \hbar\sigma_{\mathrm{z}}/2$ at the end of the evolution, for different Zeeman energies as a function of the optical pulse area and acoustic frequency. \subfigref{fig:map_evolution}{a} shows the results for $\hbar\omega_{\mathrm{Z}} = 80$~\textmu{}eV, $\hbar\Delta_{0} \approx 3.94$~meV, and $\hbar A=2$~meV. The optimal parameters for the acousto-optical $\pi$-rotation are $\varTheta\approx69.1\pi$ and $\hbar\omega_{\mathrm{ac}}\approx99.88$~\textmu{}eV. The latter corresponds to SAWs at a frequency of $24.15$~GHz. The higher maxima correspond to the $(2n-1)\pi$-rotations, for which the optimal acoustic frequency changes slightly due to the AC Stark shift. To show impact of detuning, \subfigref{fig:map_evolution}{b} shows analogous results for $\hbar\omega_{\mathrm{Z}} = 160$~\textmu{}eV, $\hbar\Delta_0\approx3.44$~meV, and $\hbar A = 2$~meV. Here, the optimal parameters change to $\varTheta\approx38.4\pi$ and $\hbar\omega_{\mathrm{ac}}\approx181.46$~\textmu{}eV ($43.88$~GHz). 

The evolution itself is shown in \subfigref{fig:map_evolution}{c}, where one can recognize the ``swing-up''-like evolution of the spin qubit (black solid and red dashed lines)~\cite{Bracht2021, Kuniej2025}, leading to an exact $\pi$ rotation (within numerical precision) over the ${\sim}155$~ps gate duration (defined as $4\sigma_{\mathrm{L}}$). This timescale is significantly shorter than the ${\sim}45$~ns acousto-optical noise-mitigating $X$ gate in Ref.~\onlinecite{KuniejPRL2026}, the ${\sim}2.2$~ns Raman $\pi$-rotation time recently demonstrated for all-optical spin control in the Faraday configuration~\cite{Koong2026, Kaspari2026}, and the typical nanosecond-scale radio-frequency or electric-dipole electron-spin-control pulse durations~\cite{Koppens2006, Nowack2007, Yoneda2014}. While the method remains slower than picosecond all-optical spin control in the Voigt configuration~\cite{Press2008}, its gate speed is competitive with established spin-control techniques while additionally introducing an acoustic interface. Additionally, the speed is limited by the acoustic frequencies that can currently be generated by interdigital transducers, so the continued development of this technology~\cite{Yaremkevich2021, Zhou2023, Priya2023} or switching to other generation techniques~\cite{Maznev2012, Ruello2015, Akimov2015} could further reduce the gate time.

The occupation of the trion state [dotted blue line, \subfigref{fig:map_evolution}{c}] remains relatively low. In particular, for the $\hbar\Delta_0\ge3$~meV regime used here, we remain within the validity range of the adiabatic elimination performed in the effective model. However, since the evolution timescale is a non-negligible fraction of the typical trion lifetimes in single QDs, its radiative decay may lead to noticeable dephasing. Thus, in \subfigref{fig:map_evolution}{d} we show that the trion occupation can be reduced via detuning. The red triangles show the maximal instantaneous trion occupation for optimized laser and acoustic parameters needed for a $\pi$-rotation as a function of $\hbar\Delta_0$. The red dashed line shows a $1/\Delta_0$ dependence. Additionally, blue squares (right axis) present the optimal laser pulse area, which scales as $\sqrt{\Delta_0}$ (blue dashed line). Both dependencies agree with the intuition given by the Hamiltonian and optical shift strength [\eqnref{eq:spinQDHamiltonian} and \eqnref{eq:staticShift}, respectively]. The maximal trion occupation decreases as $1/\Delta_0$, as expected for an off-resonantly driven $\Lambda$-system at fixed effective coupling~\cite{Economou2006, Press2008}. Moreover, as follows directly from Eq.~\eqref{eq:staticShift}, maintaining a constant optical shift while varying the detuning requires $\varTheta_{\mathrm{opt}}\sim\sqrt{\Delta_0}$.

\subsection*{Universal qubit control}

\begin{figure}[tb!]
    \centering
    \includegraphics[width=1\columnwidth]{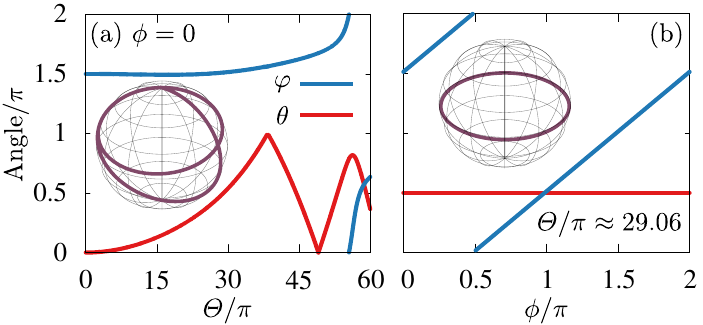}
    \caption{Position on the Bloch sphere after the pulse for an initially spin-right state, parametrized by spherical angles $\theta, \varphi$ (a) as a function of laser pulse area. Other parameters as in \subfigref{fig:map_evolution}{b}. (b) The same but for a fixed laser power ($\varTheta/\pi\approx29.06$), plotted as a function of the acoustic phase $\phi$.}
    \label{fig:Angle}
\end{figure}

From the effective model [\eqnref{eq:hamiltonianTimeIndependent}], we know that both the inclination and azimuth of the qubit rotation axis can be controlled. In particular, changing the acoustic phase
rotates the transverse component of the axis around $z$, while the detuning controls its inclination. The resulting non-collinear axes provide non-commuting rotations and hence universal qubit control.
In \figref{fig:Angle} we demonstrate the results of the control gate for a given initial state of the system ($\ket{\rightarrow}$). The final state is parametrized via spherical angles on a Bloch sphere $\ket{\mathit{\Psi}} = \cos(\theta/2)\ket{\rightarrow} + e^{i\varphi}\sin(\theta/2)\ket{\leftarrow}$.
In \subfigref{fig:Angle}{a}, we show how both angles change as a function of $\varTheta$ for the same parameters as in \subfigref{fig:map_evolution}{b}. The solid red line shows the behavior of the colatitude, which reaches $\pi$ for $\varTheta\approx38.4\pi$, as in \subfigref{fig:map_evolution}{b}. During this, the azimuth (solid blue line) also changes non-linearly. In \subfigref{fig:Angle}{b}, we vary the acoustic phase in a $\theta=\pi/2$ rotation and notice that the nonlinearity in $\varPhi_n(\phi)$ is negligible, as could be expected for $\mathcal{A}\ll1$, for which $\varPhi_n(\phi)\to n\phi$. Since the colatitude (solid red line) does not change, we are always able to prepare the desired spin superposition, while the manipulation of the phase can be used to periodically control the relative phase between spin states (solid blue line),
demonstrating independent access to the relative phase of the prepared spin superposition, as predicted through the effective model.

\begin{figure}[!t]
    \centering
    \includegraphics[width=1.\linewidth]{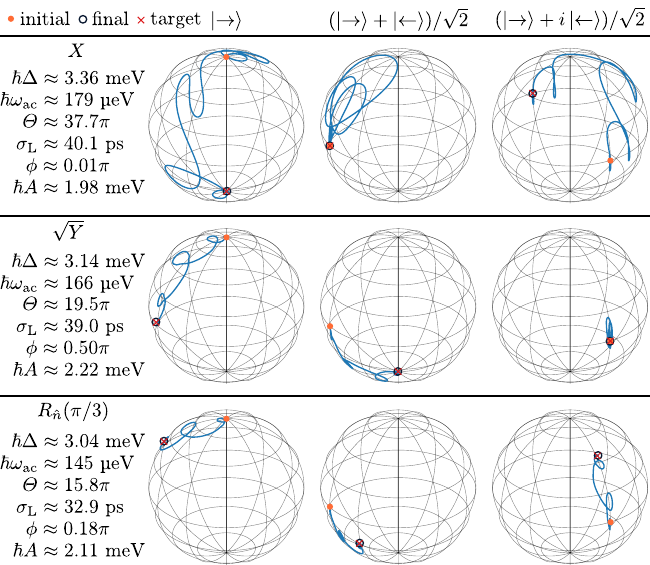}
    \caption{Demonstration of spin control in the $\omega_{\mathrm{Z}}$ rotating frame for three selected gates: $X$ (top row), $\sqrt{Y}$ (middle), and $R_{\hat{n}}(\pi/3)$ with $\hat{n}\propto(3,2,1)$ (bottom) for three exemplary initial states: $\ket{\rightarrow}$ (left column), $(\ket{\rightarrow}+\ket{\leftarrow})/\sqrt{2}$ (middle), and $(\ket{\rightarrow}+i\ket{\leftarrow})/\sqrt{2}$ (right). Filled, open, and cross symbols denote the initial, final, and target states, respectively; solid lines show qubit trajectories. In all the cases we use $\hbar\omega_{\mathrm{Z}} = 160$~\textmu{}eV. The remaining parameters are gate-specific and are listed below the gate symbol.
    }
    \label{fig:Control}
\end{figure}

Having established the possibility of arbitrary state preparation after a reset, we now turn to universal qubit control. Hamiltonian~\eqref{eq:hamiltonianTimeIndependent} predicts that nearly arbitrary rotation axes are accessible, as also illustrated in \subfigref{fig:systemQD}{e}. In \figref{fig:Control}, we numerically demonstrate the action of an $X$ gate, a $\sqrt{Y}$ gate, and a more general $\pi/3$ rotation about $\hat{n}\propto3\hat{x}+2\hat{y}+\hat{z}$ for three different initial states, $\ket{z_{+}}$, $\ket{x_{+}}$, $\ket{y_{+}}$, ($\ket{z_{\pm}}=\ket{\rightarrow\!{/}\!\leftarrow}$, $\ket{x_{\pm}}=(\ket{\rightarrow}\pm\ket{\leftarrow})/\sqrt{2}$, $\ket{y_{\pm}}=(\ket{\rightarrow}\pm i\ket{\leftarrow})/\sqrt{2}$) in the rotating frame at frequency $\omega_{\mathrm{Z}}$. Filled, open, and cross symbols denote the initial, final, and target states, respectively, while the solid lines show the exact trajectories. In all cases, the exact evolution reaches the corresponding target state (in the absence of decoherence sources, the gates have an average fidelity of unity up to our numerical precision of $10^{-6}$), confirming the predictions of the effective model and demonstrating universal qubit control.

\subsection*{Gate fidelity}
Up to this point, all simulations were performed assuming unperturbed evolution. We expect the precision and robustness of the spin control protocol to be limited by two independent factors: the limited radiative lifetime (varying from sub- and single-nanosecond values in single QDs~\cite{Dalgarno2008, Dusanowski2018, Koong2026} to hundreds of nanoseconds in QD molecules~\cite{delaGiroday2011}) and quasistatic magnetic noise.

The latter arises due to slow random evolution of $10^4$--$10^5$ nuclear spins within a QD~\cite{JacksonPRX2022}, forming the quasi-static random Overhauser field $\bm{\delta}$. The impact is described by the Hamiltonian
\begin{equation}
    H_{\delta} = \frac{1}{2}\hbar\bm{\delta}\cdot\bm{\sigma}.
\end{equation}
Under the rotating wave approximation and in the rotating frame with respect to $\omega_{\mathrm{Z}}$, the system Hamiltonian \eqref{eq:spinQDHamiltonian} reads
\begin{multline}
        \mkern-16mu H^{(\mathrm{rf})}\mkern-2mu(t) = \hbar\Delta(t)\ketbra{\mathrm{T}}{\mathrm{T}} \\
        +\frac{1}{2}\hbar\varTheta f(t)\left(e^{\frac{i}{2}\omega_{\mathrm{Z}}t}\ketbra{\rightarrow}{\mathrm{T}} + e^{-\frac{i}{2}\omega_{\mathrm{Z}}t}\ketbra{\leftarrow}{\mathrm{T}} + \hc\right)\!.\!\!\!
\end{multline}
Under these transformations, the $\sigma_{\mathrm{z}}$-component of the Overhauser noise does not change, and the transverse noise can be neglected due to its nonsecular evolution. Thus, we are left with the term
\begin{equation}
    H_{\delta} = \frac{1}{2}\hbar\delta\sigma_{\mathrm{z}},
\end{equation}
which produces an additional unknown magnetic field aligned with the applied external magnetic field, thereby modifying the Zeeman splitting and making the acoustic field slightly off-resonant. 

The random nuclear spin ensemble can be described as a zero-mean normal distribution $\mathcal{N}(\delta; 0, \sigma_{\delta})$ with standard deviation $\sigma_{\delta} = \sqrt{2}/T_{2}^{*}$, where $T_{2}^{*}$ is the spin transverse dephasing time~\cite{Cywinski2009}. A single gate realization is coherent but imprecise, and probing different noise realizations yields an average gate infidelity. We calculate the impact of both sources of infidelity by numerically solving \eqnref{eq:Lindblad}, varying $\delta$ for a given recombination time $\tau$ and averaging the final density matrix $\rho^{(\delta, \tau)}(\infty)$ over the distribution of $\delta$ for each of the six initial states $\ket{\mu}$, $\mu\in\{x_{\pm},y_{\pm},z_{\pm}\}$
\begin{equation}
    \overline{\rho}_{\mu}^{(\sigma_\delta, \tau)}(\infty) = \int_{-\infty}^{+\infty}\mathrm{d}\delta \, \mathcal{N}(\delta; 0, \sigma_{\delta})\rho_{\mu}^{(\delta, \tau)}(\infty).
\end{equation}
Then the average gate infidelity $r^{(\tau)}(\sigma_{\delta}) = 1- F^{(\tau)}(\sigma_{\delta})$ of the gate is given by projecting onto the target final states $\rho_{\mu}^{(\mathrm{target})} = U_{\mathrm{target}} \, \rho_{\mu}(0) \, U_{\mathrm{target}}^\dagger$~\cite{Bowdrey2002},
\begin{equation}
    r^{(\tau)}(\sigma_{\delta}) = 1 - \frac{1}{6}\sum_{\mu}\mathrm{Tr}\left[ \rho_{\mu}^{(\mathrm{target})}(\infty)\,\overline{\rho}_{\mu}^{(\sigma_{\delta}, \tau)}(\infty)\right].
\end{equation}

\begin{figure*}[!tb]
    \centering
    \includegraphics[width=1\linewidth]{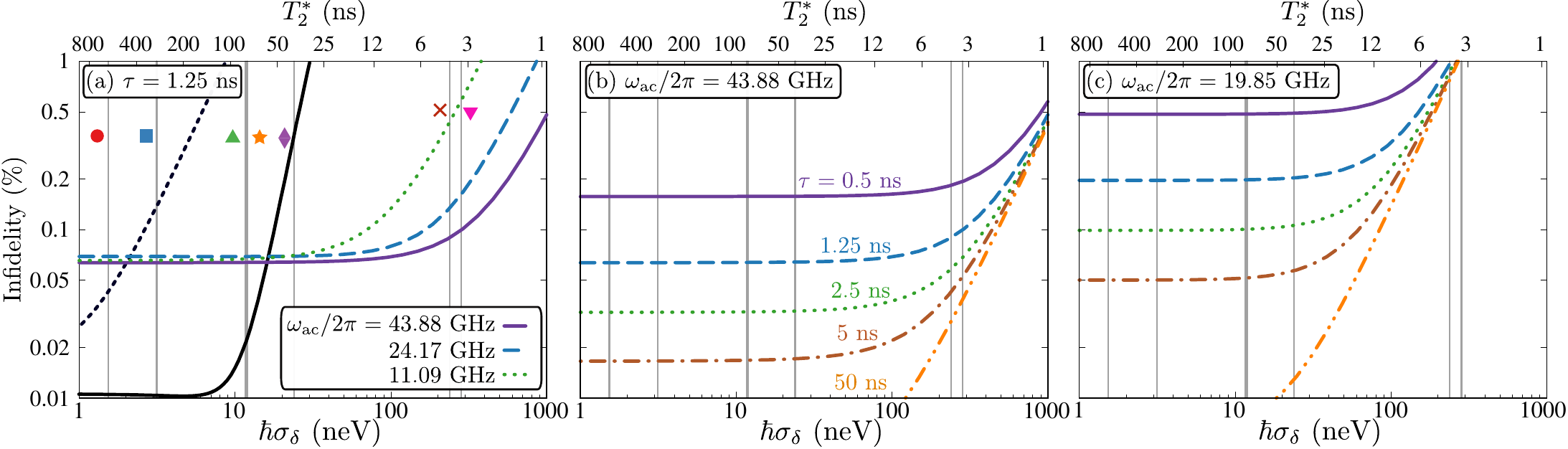}
    \caption{Average gate infidelity for the $X$ gate as a function of the Overhauser root-mean-square energy $\hbar\sigma_{\delta}$ for selected fixed trion lifetimes. Vertical lines mark $T_2^*$ times for relevant systems: \symRedCircle{} Ref.~\onlinecite{NguyenPRL2023}, \symBlueSquare{} Ref.~\onlinecite{JacksonPRX2022}, \symGreenTri{} Ref.~\onlinecite{SunPRL2012}, \symOrangeStar{} Ref.~\onlinecite{NguyenPRL2023}, \symPurpleDiam{} Ref.~\onlinecite{EthierMajcherPRL2017}, \symBrownCross{} Ref.~\onlinecite{NguyenPRL2023}, \symPinkDownTri{} Ref.~\onlinecite{JacksonPRX2022}. Parameters used in simulations are given in Table~\ref{tab:parameters}. (a) Fixed $\tau = 1.25$~ns and different values of $\omega_{\mathrm{ac}}/2\pi$: $43.88$~GHz (solid purple line), $24.17$~GHz (dashed blue), $11.09$~GHz (dotted green). (b) Fixed $\omega_{\mathrm{ac}}/2\pi = 43.88$~GHz and different values of $\tau$: $0.5$~ns (solid purple line), $1.25$~ns (blue long-dashed), $2.5$~ns (green dotted), $5$~ns (brown dashed-dotted), and $50$~ns (orange dotted-dotted-dashed). Dotted and solid black lines show the results from Ref.~\onlinecite{KuniejPRL2026} and correspond to the two-pulse naive gate and a noise-mitigating sequence, respectively. (c) The same as (b) but for an acoustic field tuned to two-phonon processes with $\omega_{\mathrm{ac}}/2\pi \approx 19.85$~GHz.}
    \label{fig:decoherence}
\end{figure*}

\begin{ruledtabular}
\begin{table}[tb!]
    \centering
    \small
    \begin{tabular}{lcccc}
         & 11.09 GHz & 24.17 GHz & 43.88 GHz & Two-phonon \\[2pt]
        \hline\vspace{2px}
        $\hbar\omega_{\mathrm{Z}}$~(\textmu{}eV) & 40 & 80 & 160 & 160 \\\vspace{2px}
        $\hbar\Delta_0$~(meV) & 3.475 & 3.943 & 3.448 & 3.000\\\vspace{2px}
        $\hbar A$~(meV) & 1.999 & 2.001 & 2.002 & 2.001\\\vspace{2px}
        $\sigma_{\mathrm{L}}$~(ps) & 154.6 & 78.64 & 38.72 & 207.1\\\vspace{2px}
        $\varTheta/\pi$ & 75.48 & 69.21 & 38.37 & 116.9\\\vspace{2px}
        $\phi/\pi$ & 1.000 & 1.000 & 0.000 & 0.500 \\
    \end{tabular}
    \caption{Parameters used in simulations for \figref{fig:decoherence}.}\label{tab:parameters}
\end{table}
\end{ruledtabular}

In \figref{fig:decoherence}, we show $r^{(\tau)}(\sigma_{\delta})$ as a function of $\hbar\sigma_{\delta}$ and $T_2^*$ (top axis) for exemplary evolutions implementing the $X$ gate with parameters used in simulations shown in Table~\ref{tab:parameters}. We use the $X$ gate as a representative conservative reference. The acoustic phase only sets the gate-axis azimuth, while both the longitudinal Overhauser noise and the trion decay are invariant under $z$-axis rotations, so the calculated infidelity is identical for $\pi$ rotations about other transverse axes (e.g., $Y$). Moreover, both error contributions increase with gate time, so shorter-angle rotations are expected to have higher fidelity. \asubfigref{fig:decoherence}{a} shows the results for $\tau = 1.25$~ns. The vertical lines show the literature $T_2^*$ values for various spin-cooling techniques and an unprepared thermal environment. We show results for three different acoustic-field frequencies (and therefore different Zeeman energy and evolution timescale): $11.09$~GHz (dotted line), $24.17$~GHz (dashed), and $43.88$~GHz (solid). The infidelity remains below $0.1\%$ for all noise amplitudes corresponding to cooled environments and even thermal environments in both GaAs and InAs QDs, if the gate time is short enough (solid line). Even for longer gates, $r^{(\tau)}(\sigma_{\delta})$ is low, and stays below $0.8\%$ for all considered cases. Additionally, we compare those infidelities with results from Ref.~\onlinecite{KuniejPRL2026}: black lines calculated for $\omega_{\mathrm{ac}}/2\pi = 44$~GHz for naive two-pulse gate implementation (dashed) and noise cancellation sequence (solid) for infinite trion lifetime. Since our protocol is ${\sim}290\times$ faster than the noise-mitigating gate (or ${\sim}160\times$ compared to the naive implementation, offering incomparably lower fidelity), the accumulated error due to the Overhauser field remains low compared to the previous work. Even without introducing a noise cancellation sequence to the proposed protocol, the infidelity stays below $0.1\%$ for ${\sim}$20$\times$ shorter $T_2^*$ times, marking progress from proof-of-principle to a potentially practically usable method.

In \subfigref{fig:decoherence}{b}, we study how the trion lifetime in the range of 0.5--50~ns affects the scheme. We focus on a single frequency of the acoustic field (43.88~GHz) corresponding to the shortest evolution. The region below $\hbar\sigma_{\delta}\approx100$~neV is purely dominated by the recombination of the trion, which limits the fidelity of the gate when $T_2^*$ is large enough. For an extended trion lifetime (e.g., for an orbitally dark trion, or an indirect trion in a double QD), its influence is negligible, and the remaining infidelity comes only from the fluctuations of the Overhauser field, which become meaningful only for the thermal environment, and even in this regime, $r^{(\tau)}(\sigma_{\delta})$ remains below $0.1\%$ for $\tau\gtrsim1.25$~ns, which shows the method should provide high fidelity in real QD platforms.

In \subfigref{fig:decoherence}{c}, we show the average infidelity of the same $X$ gate realized using second harmonic acoustic resonance. As shown in Table~\ref{tab:parameters}, the required evolution time is in all first-harmonic cases shorter than in two-phonon parametric resonance. Thus, the latter is more strongly limited by both the finite trion lifetime and the Overhauser field fluctuations. However, we still predict high fidelity for the higher-harmonic control gate, for all considered cooled environments when $\tau \gtrsim 2.5$~ns. This approach may become particularly useful for controlling and entangling multi-phonon states in acoustic resonators coupled to a charged QD.

\section*{Discussion}
We have proposed a parametric control method for a quantum-dot spin qubit based on the resonance between acoustic modulation and the optically dressed spin splitting. The method removes both the speed and restricted-rotation-axis bottlenecks of the previous proof-of-concept method~\cite{KuniejPRL2026} and provides universal spin control, with a nearly arbitrary spin-rotation axis determined by the acoustic phase and detuning from the parametric resonance. The simulated ${\sim}155$~ps gate time is more than two orders of magnitude shorter than the proof-of-concept protocol and over an order of magnitude shorter than for all-optical Faraday control and far exceeds conventional electric-dipole methods. This fast operation suppresses Overhauser-noise errors even without noise-mitigating sequences and limits the impact of trion recombination. 

Although it requires relatively high acoustic frequencies and modulation amplitudes ($43.88$~GHz and $2$~meV, corresponding to ${\sim}2$--$3\times 10^{-4}$ strain and few-picometer displacements), the approach can provide average gate fidelity above 99.9\% even for an uncooled nuclear spin environment and with a realistic ${\sim}1.25$~ns trion lifetime, the latter of which limits fidelity in the low-noise regime. For cooled nuclear environments, significantly lower acoustic frequencies allow reaching the 99.9\% fidelity threshold.
Thus, the method provides fast, universal spin manipulation that is competitive with current methods while potentially introducing the spin-phonon interface into the system, including access to multi-phonon excitations via available higher-harmonic resonances.
Our effective model not only predicts the resonance conditions and accessible rotation axes but also reveals the analogy to the ``swing-up'' excitation method for charge states in quantum emitters~\cite{Bracht2021}.

Acoustic spin rotation also has an important integration advantage due to the low sound velocity, which yields a shorter wavelength at the given frequency, enabling compact on-chip structures with the possibility of in situ generation using interdigital transducers.
With established interfaces to light, microwaves, nuclear spins, and phonons, QDs could become fundamental elements in hybrid quantum systems, potentially capable of acoustic state transfer and controllable spin-phonon entanglement.

\appendix
\renewcommand{\theequation}{A\arabic{equation}}
\setcounter{equation}{0}
\section*{Appendix A: Three-level description in the Voigt geometry}\label{app:three-level}
A negatively charged QD in the Voigt geometry generally forms a four-level double-$\Lambda$ system with two electron spin states and two trion states~\cite{SchimpfPRX2025}. Here, we show that under far-detuned circularly polarized excitation, this structure effectively reduces to the three-level model used in the main text.

Let $\ket{\mathrm{T}_{+}}$ be the trion state that can be exicted with a $\sigma^{+}\!$-polarized light from the electron-spin $\ket{\uparrow}$ state (and analogously $\ket{\mathrm{T}_{-}}$ from $\ket{\downarrow}$ with $\sigma^{-}$). In the Voigt geometry, the electron eigenstates are $\ket{\rightarrow}=(\ket{\uparrow}+\ket{\downarrow})/\sqrt{2}$ and $\ket{\leftarrow}={(\ket{\uparrow}-\ket{\downarrow})/\sqrt{2}}$ and are both equally coupled to the $\ket{\mathrm{T}_{+}}$ state by a $\sigma^{+}$ light, while $\ket{\mathrm{T}_{-}}$ remains not optically addressed under such driving. Taking into account the Zeeman effect for the trion, the complete four-level Hamiltonian in the rotating frame at the laser frequency is
\begin{align}\label{eq:app-4lvl}
    H_{4\mathrm{lvl}}(t) = {}&
        \frac{1}{2}\hbar \omega_{\mathrm{Z}} \lr*{ \ketbra{\rightarrow}{\rightarrow} - \ketbra{\leftarrow}{\leftarrow} } \notag \\
        {}& + \frac{1}{2}\hbar\omega_{\mathrm{h}} \lr*{ \ketbra{\mathrm{T}_{+}}{\mathrm{T}_{-}} + \hc } \notag \\
        {}& + \hbar\Delta(t) \lr*{ \ketbra{\mathrm{T}_{+}}{\mathrm{T}_{+}} + \ketbra{\mathrm{T}_{-}}{\mathrm{T}_{-}} } \notag \\
        {}& +\frac{1}{2}\hbar\varTheta f(t) \Lr*{ \lr*{ \ket{\rightarrow} + \ket{\leftarrow} }\bra{\mathrm{T}_{+}} +\hc  },
\end{align}
where $\omega_{\mathrm{h}}$ is the trion transverse Zeeman splitting, which for heavy-hole trions arises only from perturbative effects and is thus typically low and anisotropic, which allows for additional minimization~\cite{Semenov2003, Ramesh2025, SchimpfPRX2025}. The acoustic coupling is spin-independent, so both trion states are modulated identically.

For $\omega_{\mathrm{h}}=0$, $\ket{\mathrm{T}_{-}}$ state is exactly decoupled, and \eqnref{eq:app-4lvl} separates into the energy of that state and the three-level Hamiltonian from \eqnref{eq:spinQDHamiltonian}. For nonzero $\omega_{\mathrm{h}}$, the $\ket{\mathrm{T}_{-}}$ is still not addressed optically, but can become indirectly populated via the Zeeman term. Let the system state be
\begin{equation}
    \ket{\varPsi(t)} = c_{\rightarrow}(t)\ket{\rightarrow} + c_{\leftarrow}(t)\ket{\leftarrow} + c_{+}(t)\ket{\mathrm{T}_{+}} + c_{-}(t)\ket{\mathrm{T}_{-}},
\end{equation}
for which, through the Schr\"odinger equation, we get for the $c_{-}$ amplitude
\begin{equation}
    i\dot{c}_{-}(t) = \Delta(t) c_{-}(t) + \frac{1}{2}\omega_{\mathrm{h}}c_{+}(t).
\end{equation}
Under far-detuned optical field, we have $\Delta_{\mathrm{min}}=\abs{\Delta_0}-A \gg \abs{\omega_{\mathrm{h}}},\abs{\dot{c}_{+}}$, and $c_{-}(t)$ adiabatically follows $c_{+}(t)$,
\begin{equation}
  c_{-}(t) \simeq - \frac{\omega_{\mathrm{h}}}{2\Delta(t)}c_{+}(t).
\end{equation}
Thus, the occupation of $\ket{\mathrm{T}_{-}}$ remains perturbatively small $p_{\mathrm{T}_{-}}\!\sim\abs{\omega_{\mathrm{h}}/2\Delta_{\mathrm{min}}}^2p_{\mathrm{T}_{+}}$. Its elimination only renormalizes the detuning,
\begin{equation}
    \Delta'(t) \simeq \Delta(t)-\frac{\omega_{\mathrm{h}}^2}{4\Delta(t)},
\end{equation}
while reproducing the Hamiltonian from \eqnref{eq:spinQDHamiltonian}. Taking the literature ratios of transverse trion and electron $g$-factors $g_{\mathrm{T}}/g_{\mathrm{e}}$ of $0.65$ for InAs~\cite{Xu2007} and $1.18$ for GaAs~\cite{NguyenPRL2023}, the estimated maximal $p_{\mathrm{T}_{-}}$ remains below  $0.003\,p_{\mathrm{T}_{+}}$ and $0.009\,p_{\mathrm{T}_{+}}$, respectively, for all parameter sets from Table~\ref{tab:parameters}, while $p_{\mathrm{T}_{+}}$ itself is kept perturbatively low.
Direct numerical checks show relative infidelity differences of less than $2\%$ (${\sim}0.001\%$ absolute differences) between the three- and four-level models (interestingly, with the four-level model often exhibiting higher fidelity) for $g_{\mathrm{T}}/g_{\mathrm{e}}$ in the range up to 5, after a narrow-range re-optimization that primarily compensates for the aforementioned detuning renormalization.

\renewcommand{\theequation}{B\arabic{equation}}
\setcounter{equation}{0}
\section*{Appendix B: Nonperturbative treatment of the acoustic modulation}\label{app:nonperturbative}
In the main text, we presented a simple and transparent derivation of the effective Hamiltonian by expanding the modulated detuning to leading order in the modulation depth $a=A/\Delta_0$. Here, we present a more rigorous derivation that is nonperturbative in $a$ and thus accounts for all nonlinearities leading to a dressed spin splitting renormalization and retains higher harmonics in both longitudinal and transverse modulations, as well as their interplay. We mark quantities that are nonperturbative equivalents of those in the main text with a prime. Finally, we show that the rigorous result reproduces the leading-order theory from the main text in the $a\ll1$ limit, while beyond that limit it has the same structure but with renormalized qubit rotation-axis parameters.

We begin with the Hamiltonian from \eqnref{eq:eff-H-initial}
\begin{equation}
    H_{\mathrm{eff}}(t) = \frac{1}{2}\hbar\omega_{\mathrm{Z}}\sigma_{\mathrm{z}} - \frac{1}{2} \hbar\omega_{\mathrm{s}}(t) \frac{1}{1+a\cos\lr*{\omega_{\mathrm{ac}}t+\phi}}\sigma_{\mathrm{x}}.
\end{equation}
To isolate the purely optical part of the Hamiltonian $H_{\mathrm{eff}}^{(0)}(t) = H_{\mathrm{eff}}(t)\big|_{A=0}$, we rewrite
\begin{equation}
    \frac{1}{1+a\cos\lr*{\omega_{\mathrm{ac}}t+\phi}} = 1-\eta\lr*{t},
\end{equation}
with
\begin{equation}
    \eta(t) = \frac{a\cos\lr*{\omega_{\mathrm{ac}}t+\phi}}{1+a\cos\lr*{\omega_{\mathrm{ac}}t+\phi}},
\end{equation}
and obtain the nonperturbative acousto-optical part
\begin{equation}
    H_{\mathrm{eff}}^{\prime(\mathrm{ac})}\!(t) = \frac{1}{2}\hbar\omega_{\mathrm{s}}(t) \eta(t) \, \sigma_{\mathrm{x}}.
\end{equation}
In the optically-dressed state basis from \eqnref{eqs:dressedStates}, this reads
\begin{equation}
    \widetilde{H}_{\mathrm{eff}}^{\prime(\mathrm{ac})}\!(t) = \frac{1}{2}\hbar \Lr*{ h_z(t) \, \widetilde{\sigma}_{\mathrm{z}}  + h_x(t) \, \widetilde{\sigma}_{\mathrm{x}} },
\end{equation}
where the longitudinal and transverse modulations
\begin{align}
    h_x(t) = {}& \frac{\omega_{\mathrm{s}}(t) \omega_{\mathrm{Z}} }{\SpinSplit(t)} \eta(t), \\
    h_z(t) = {}& - \frac{\omega_{\mathrm{s}}^2(t) }{\SpinSplit(t)} \eta(t),
\end{align} 
both contain the same nonlinearity through $\eta(t)$ but differ in the amplitude, $h_x(t) = -(\omega_{\mathrm{Z}}/\omega_{\mathrm{s}}) h_z(t)$.
The nonoscillating part of $\eta(t)$,
\begin{equation}
    \eta_0=\frac{\omega_{\mathrm{ac}}}{2\pi}\int_0^{{2\pi}/{\omega_{\mathrm{ac}}}} \!\!\mathrm{d}t \, \eta(t) = 1 - \frac{1}{\sqrt{1-a^2}},
\end{equation}
reveals the renormalization of the dressed spin splitting
\begin{equation}
    \SpinSplit'(t)=\SpinSplit(t) - \frac{\omega_{\mathrm{s}}^2(t) }{\SpinSplit(t)} \eta_0.
\end{equation}

As in the leading-order derivation, we consider the plateau of a rectangular-like optical pulse and treat $\omega_s$ and $\Omega$ as constant over an acoustic period. We remove the oscillatory part $h_z^{(\mathrm{osc})}(t)=h_z(t)+{\omega_{\mathrm{s}}^2(t) }/{\SpinSplit(t)} \eta_0$ from the diagonal via the unitary transformation
\begin{equation}
    S'(t) = \frac{1}{2} \int_0^t \! \mathrm{d}\tau \, h_z^{(\mathrm{osc})}(\tau) \, \widetilde{\sigma}_z = \frac{1}{2} \Lr*{ F(t) - F(0) } \, \widetilde{\sigma}_z,
\end{equation}
where $\mathrm{d}F(t)/\mathrm{d}t =  h_z^{(\mathrm{osc})}(t)$. This yields
\begin{align}\label{eq:app-eff-H-tilde}
    \widetilde{\mathcal{H}}_{\mathrm{eff}}'(t) =
        \frac{1}{2}\hbar\SpinSplit'(t) \, \widetilde{\sigma}_z +  \frac{1}{2} \hbar \Lr*{ e^{-iF(0)}g(t)  \, \widetilde{\sigma}_{+} + \hc },
\end{align}
where $g(t) = h_x(t) e^{iF(t)}$ is periodic in $t$ and thus also in the angle $\xi=\omega_{\mathrm{ac}}t+\phi$. We thus expand it into a Fourier series
\begin{equation}
    g(t) = \sum_{n=-\infty}^{\infty} \lr*{ 2g_n^{\prime} } e^{-in(\omega_{\mathrm{ac}}t+\phi)},
\end{equation}
where
\begin{equation}
    g_n^{\prime}\! = \frac{1}{4\pi} \int_0^{2\pi} \! \mathrm{d}\xi \, g(\xi) e^{in\xi},
\end{equation}
and we chose the integration constant of $F(t)$ such that $g_n^{\prime}$ are real. By inserting this into \eqnref{eq:app-eff-H-tilde}, switching to the $n\omega_{\mathrm{ac}}$ rotating frame, and performing the secular approximation, we arrive at a time-independent effective Hamiltonian
\begin{align}\label{eq:app-H-eff-final}
    \widetilde{\mathcal{H}}_{\mathrm{eff},n}^{\prime(\mathrm{rf})}(t) = {}& \frac{1}{2}\hbar\delta_n^{\prime} \, \widetilde{\sigma}_z
        + \hbar g_n^{\prime} \lr*{ e^{-i\varPhi_n^{\prime}(\phi)} \,  \widetilde{\sigma}_{+} + \hc } \notag \\
        = {}& \frac{1}{2}\hbar \bm{\varOmega}_{n}^{\prime}\!\cdot \widetilde{\bm{\sigma}}
\end{align}
with the rotation axis
\begin{equation}
    \bm{\varOmega}_{n}^{\prime} = \lr*{ 2g_n^{\prime} \! \cos\varPhi_n^{\prime}(\phi), 2g_n^{\prime} \! \sin\varPhi_n^{\prime}(\phi), \delta_n^{\prime}},
\end{equation}
where $\varPhi_n^{\prime}(\phi) = n\phi + F(0)$, and $ \delta_n^{\prime} = \varOmega' - n\omega_{\mathrm{ac}}$.

The structure of Hamiltonian~\eqref{eq:app-H-eff-final} is identical to that of \eqnref{eq:hamiltonianTimeIndependent}. The nonperturbative treatment of nonlinearities renormalized the rotation vector through the change of the dressed spin splitting $\varOmega'$ and consequently detuning $\delta_n^{\prime}$ as well as the drive strength $g_n^{\prime}$ and the azimuthal phase $\varPhi_n^{\prime}(\phi)$. In the limit of low modulation depth, we recover
\begin{subequations}
\begin{align}
    \delta_n^{\prime} \xrightarrow{a\ll1} {}& \delta_n, \\
    g_n^{\prime} \xrightarrow{a\ll1} {}& g_n, \\
    \varPhi_n^{\prime}(\phi) \xrightarrow{a\ll1} {}& \varPhi_n(\phi) ,
\end{align}
\end{subequations}
and thus $ \bm{\varOmega}_{n}^{\prime} \!\xrightarrow{a\ll1} \bm{\varOmega}_{n}$, which exactly reproduces the leading-order effective model given in \eqnref{eq:hamiltonianTimeIndependent} of the main text.
\renewcommand{\theequation}{C\arabic{equation}}
\setcounter{equation}{0}
\section*{Appendix C: Details of main-text derivations}\label{app:detailsOfTransformation}
Our goal in deriving the effective model is to transform the effective Hamiltonian to the form of the standard Rabi model. As shown in the main text, the Hamiltonian from \eqnref{eq:effectiveAcousticDressedStates} has a time-dependent $\widetilde{\sigma}_{\mathrm{z}}$ component that needs to be transformed out. We achieve this and obtain the Hamiltonian \eqref{eq:effectiveHamiltonian} by applying the unitary transformation from \eqnref{eq:unitaryTransformation}, i.e., $U_{S}(t) = e^{iS(t)}$. Its action on $\widetilde{\sigma}_{x}$ is
\begin{equation}
    U_{S}(t)\widetilde{\sigma}_{\mathrm{x}}U^{\dagger}_{S}(t) \!=\!\!\!\! \sum_{m=-\infty}^{+\infty}\!\!\!J_m(\mathcal{A})e^{-i\Lr*{m\omega_{\mathrm{ac}}t + \varPhi_m(\phi)}}\ketbra{+}{-} + \hc,
\end{equation}
where we also applied the Jacobi-Anger expansion. In the Hamiltonian, $\widetilde{\sigma}_{x}$ is additionally multiplied by $\omega_{\mathrm{d}}(t)$, and thus yet another frequency multiplies the above series. We can group these additional frequencies with those already existing in the expansion by shifting the summation index and applying the identity $J_{m+1}(a)+J_{m-1}(a) = 2mJ_{m}(a)/a$. This yields
\begin{multline}
    U_{S}(t)\cos(\omega_{\mathrm{ac}}t + \phi)\widetilde{\sigma}_{\mathrm{x}}U^{\dagger}_{S}(t) \\= \sum_{m=-\infty}^{+\infty}\frac{m}{\mathcal{A}}J_{m}(\mathcal{A})e^{-i(m\omega_{\mathrm{ac}}t + \varPhi_m(\phi))}\ketbra{+}{-} + \hc
\end{multline}
In the last step, we perform a secular approximation to retain only the resonant $m=n$ phonon processes, yielding the effective transverse-coupling Hamiltonian in \eqnref{eq:effectiveHamiltonian}.

\section*{Data availability}
The datasets generated during this study and used to produce the figures are publicly available in RepOD~\cite{RepOD}.

\acknowledgments
We thank Pawe{\l} Machnikowski, Karolina S{\l}owik, and Piotr G{\l}adysz for helpful discussions during the early stage of this project. We acknowledge support from the National Science Centre (Poland) under Grants Nos.\ 2023/49/N/ST3/03931 (M.\,K.) and 2024/06/Y/ST3/00199 (M.\,G.).

\bibliography{spinSwingUp}

\end{document}